\documentclass[aip,jcp,amsmath,amssymb,reprint]{revtex4-1}

\usepackage{graphicx}
\usepackage{graphicx,color}% Include figure files
\usepackage{dcolumn}% Align table columns on decimal point
\usepackage{bm}% bold math
\begin{document}

%\newcommand{\ktheta}{k_{\theta}}
%\newcommand{\psq}{\frac{\ktheta}{2} \theta^2}
%\newcommand{\potsq}{$U_{stiff}(\theta) = \psq$}
%\newcommand{\pcos}{\ktheta (1-\cos\theta)}
%\newcommand{\potcos}{$U_{stiff}(\theta) = \pcos$}
%\newcommand{\ptanh}{\frac{\ktheta}{2} \theta^2 \left[ \tanh \left( \frac{2 E_s}{\ktheta \theta^2} \right)^{10} \right]^{1/10}}
%\newcommand{\pottanh}{$U_{stiff}(\theta) = \ptanh$}
%\newcommand{\leftexp}[2]{{\vphantom{#2}}^{#1}{#2}}
%\newcommand{\newtext}[1]{\textcolor{red}{#1}}
%\newcommand{\phiiv}{\phi_L^{iv}(t)}
%\newcommand{\phiid}{\phi_L^{id}(t)}
%\definecolor{grey}{rgb}{0.502,0.502,0.502}
%\definecolor{orange}{rgb}{1.,0.5,0.}
%\definecolor{brown}{rgb}{0.55,0.27,0.08}
%\definecolor{dgl}{rgb}{0.,0.3922,0.}

%\preprint{APS/123-QED}
\preprint{AIP/123-QED}

%\title{Manuscript Title:\\with Forced Linebreak}% Force line breaks with \\
\title{Viscoelasticity of model interphase chromosomes}

\author{Manon Valet}
%\email{}
\affiliation{
Ecole Normale Sup\'erieure de Lyon, 46 All\'ee d'Italie, 69634 Lyon Cedex 07 (France)
}

\author{Angelo Rosa}
\email{anrosa@sissa.it}
\affiliation{
Sissa (Scuola Internazionale Superiore di Studi Avanzati), Via Bonomea 265, 34136 Trieste (Italy)
}

\date{\today}% It is always \today, today,

\begin{abstract}
We investigated the viscoelastic response of model interphase chromosomes by tracking the three-dimensional motion of hundreds of dispersed Brownian particles of sizes ranging from the thickness of the chromatin fiber up to slightly above the mesh size of the chromatin solution.
In agreement with previous computational studies on polymer solutions and melts, we found that the large-time behaviour of diffusion coefficient and the experienced viscosity of moving particles as functions of particle size deviate from the traditional Stokes-Einstein relation,
and agree with a recent scaling theory of diffusion of non-sticky particles in polymer solutions.
Interestingly, we found that at short times large particles are temporary ``caged'' by chromatin spatial constraints, which thus form effective domains whose size match remarkably well recent experimental results for micro-tracers inside interphase nuclei.
Finally, by employing a known mathematical relation between the time mean-square displacement of tracked particles and the complex shear modulus of the surrounding solution,
we calculated the elastic and viscous moduli of interphase chromosomes.
\end{abstract}

\maketitle

\section{Introduction}\label{sec:Intro}

In eukaryotic cells, the nucleus is a well recognizable organelle,
which plays the role of maintaining the genome physically separated from the rest of the cell.
Inside the nucleus, the genome is organized into single bodies, the chromosomes, and
each chromosome is constituted of a variably-long linear filament of DNA and protein complexes, known as the chromatin fiber~\cite{alberts}.
In human cells, the nucleus is approximately $10$ micron wide and the length of a chromatin filament associated to a single chromosome is of the order of $1$ millimeter,
{\it i.e.} $\approx 100$ times longer.
Hence, chromatin fibers form an intricated polymer-like network inside the nucleus of the cell~\cite{maeshimaCellRep2012}.
In spite of this ``intricacy'' though, macromolecular complexes and enzymes which need to run and bind to specific
target sequences along the genome are relatively mobile inside the nucleus~\cite{maeshimaCellRep2012}.

In order to study quantitatively the dynamic properties of macromolecular compounds inside the chromatin mesh,
passive microrheology has been recently introduced~\cite{TsengWirtz2004,WirtzReview}.
Artificially-designed beads of sub-micron size are carefully injected inside the nucleus,
and their thermally-driven Brownian motion is tracked by fluorescence microscopy.
From the analysis of the microscopic passive dynamics of the beads inside the nuclear medium,
it is then possible~\cite{WirtzReview} to extract quantitative information on the viscoelastic properties of the medium.

Compared to standard rheology, microrheology offers several advantages.
Specifically, because of the feasibility to design trackable particles of sizes ranging from a few nanometers~\cite{guigas_biophysj2007_probing}
to hundreds of nanometers~\cite{TsengWirtz2004} and microns~\cite{Shivashankar_PlosOne2012}
microrheology can probe very efficiently a remarkably wide range of length- and time-scales.
This offers an unprecedented possibility to address specific questions in complex materials which can not be answered by traditional bulk rheology.
Nowadays, micro-rheology is extensively used to study biological materials
because it offers methods which, being minimally invasive, can be used to perform experiments {\it in vivo} by employing very small samples~\cite{CicutaReview}.

In this work, and along the lines of previous computational investigations
aiming at measuring the viscoelastic properties of polymer solutions and melts~\cite{KalathiGrest_PRL2014,KuhnholdPaul_PRE2014},
we have employed molecular dynamics computer simulations in order to study the diffusive behavior of (sub-)micron sized, {\it non-sticky} particles
probing the rheological properties of a coarse-grained polymer model of interphase chromosomes~\cite{RosaPLOS2008,RosaBJ2010,RosaEveraersPRL2014}.

Our approach complements and extends in many different ways the above-mentioned experimental work.
First,
passive {\it non-sticky} particles undergoing simple Brownian motion represent the simplest minimally-invasive tools whose microscopic dynamics can be directly linked to the viscoelastic properties of the surrounding medium.
Contrarily, if particles become sticky or they become actively driven as a consequence of some cellular process,
the corresponding link to the viscoelastic properties of the medium is much less transparent and more interpretative tools are needed~\cite{gal2013particle}.
Second,
by our computational approach we are able to single out the nominal contribution of chromatin fibers to the whole nuclear viscoelasticity.
We believe this also to be an important point,
as the viscoelastic response obtained through wet-lab experiments likely originates from the unavoidable coupling of chromatin fibers
with any other organelle present in the nucleus.
Third,
while the sizes of probing particles available in experiments appear to be limited~\cite{guigas_biophysj2007_probing,TsengWirtz2004,Shivashankar_PlosOne2012},
here we monitor systematically quite an extensive range of spatial scales,
from the nominal chromatin thickness up to just above the mesh (entanglement) size of the chromatin solution.

In qualitative agreement with recent findings from a computational study on entangled polymer melts~\cite{KalathiGrest_PRL2014},
we found that the diffusive behavior of probe particles as a function of particle size deviates from the traditional Stokes-Einstein picture.
Our results can be well understood instead in terms of a recently proposed~\cite{CaiPanuykovRubinstein2011} scaling theory of diffusion of non-sticky particles in polymer solutions.
We demonstrated further, that large particles at short time scales are temporary ``caged'' by chromatin spatial constraints,
and remain confined to domains whose size match remarkably well recent experimental results for micro-tracers inside interphase nuclei.
Finally, we calculated the elastic and viscous moduli of interphase chromosomes and found that,
in the available frequency range, they are more liquid- than solid-like.

The paper is organized as follows:
In Sec.~\ref{sec:fiberModel}, we describe the technical details of the chromosome polymer model,
and the theoretical framework to study the visco-elastic properties of chromatin solution.
In Sec.~\ref{sec:Results}, we present our results.
Finally, we conclude (Sec.~\ref{sec:DiscConcls}) with a brief discussion and possible perspectives about future work.

\section{Model \& methods}\label{sec:fiberModel}

\subsection{Simulation protocol I. Force field}\label{sec:MDForceField}

In this work,
the monomers used in the coarse representation of the chromatin fiber building the model chromosome
and the non-sticky micro-probes
were modeled as spherical particles, interacting through the following force field.

The intra-polymer interaction energy is the same as the one used in our previous works
on the modeling of interphase chromosomes~\cite{RosaPLOS2008,RosaBJ2010,DiStefanoRosa2013}.
It consists of the following terms:
\begin{eqnarray}\label{eq:IntraChainEnergy}
{\cal H}_{intra} & = & \sum_{i=1}^N [ U_{FENE}(i, i+1) + U_{br}(i, i+1, i+2) \nonumber\\
                        &    & + \sum_{j=i+1}^N U_{LJ}(i,j) ]
\end{eqnarray}
where $N = 39154$ is the total number of monomers constituting the ring polymer modeling the chromosome (Sec.~\ref{sec:IniConfig}), and $i$ and
$j$ run over the indices of the monomers. The latter are assumed to be
numbered consecutively along the ring from one chosen reference monomer.
The modulo-$N$ indexing is implicitly assumed because of the ring periodicity.  

By taking the nominal monomer diameter, $\sigma = 30 \mbox{ nm} = 3000 \mbox{ bp}$~\cite{RosaBJ2010},
the vector position of the $i$th monomer, $\vec{r}_i$,
the pairwise vector distance between monomers $i$ and $j$, $\vec{d}_{i,j} = \vec{r}_j - \vec{r}_i$,
and its norm, $d_{i,j}$,
the energy terms in Eq. \ref{eq:IntraChainEnergy} are given by the following expressions:

\noindent
1) The chain connectivity term, $U_{FENE}(i,i+1)$ is expressed as:
\begin{equation}\label{eq:fenepot}
U_{FENE}(i,i+1) = \left\{
\begin{array}{l}
- {k \over 2} \, R^2_0 \, \ln \left[ 1 - \left( {d_{i,i+1} \over R_0} \right)^2 \right], \, d_{i,i+1} \leq R_0\\
0, \, d_{i,i+1} > R_0
\end{array}
\right.
\end{equation}
where $R_0=1.5 \sigma$, $k=30.0 \epsilon / \sigma^2$ and
the thermal energy $k_B\, T$ equals $1.0 \epsilon$~\cite{KremerGrestJCP1990}.

\noindent
2) The bending energy has the standard Kratky-Porod form (discretized worm-like chain):
\begin{equation}\label{eq:stiffpot}
U_{br}(i, i+1 ,i+2) = \frac{k_B \, T \, \xi_p}{\sigma}  \left (1 - \frac{{\vec d}_{i,i+1} \cdot  {\vec d}_{i+1,i+2}}{d_{i,i+1}
 \, d_{i+1,i+2}} \right )
\end{equation}
where $\xi_p = 5.0 \sigma = 150 \mbox{ nm}$ is the nominal persistence length~\cite{bystricky} of the chromatin fiber.
We remind the reader, that this is equivalent to a Kuhn length, $l_K = 2 \xi_p = 300 \mbox{ mn}$~\cite{RosaPLOS2008}.

\noindent
3) The excluded volume interaction between distinct monomers (including consecutive ones)
corresponds to a purely repulsive Lennard-Jones potential:
\begin{equation}\label{eq:ljpot}
U_{LJ}(i,j) = \left\{
\begin{array}{l}
4 \epsilon [(\sigma/d_{i,j})^{12} - (\sigma/d_{i,j})^6 + 1/4], d_{i,j} \leq \sigma 2^{1/6}\\
0, d_{ij} > \sigma 2^{1/6}
\end{array}
\right. .
\end{equation}

We model the monomer-particle ($U_{mp}$) and the particle-particle ($U_{pp}$) interactions by the potential energy functions
introduced by Everaers and Ejtehadi~\cite{EveraersEjtehadi2003} for studies on colloids.

$U_{mp}$ is given by:
\begin{widetext}
\begin{equation}\label{eq:Umc}
U_{mp}(i,j) = \left\{
\begin{array}{l}
\frac{2 R^3 \sigma^3 A_{mp}}{9 (R^2-d_{i,j}^2)^3} \left[ 1 - \frac{(5R^6 + 45 R^4 d_{i,j}^2 + 63 R^2 d_{i,j}^4 + 15 d_{i,j}^6)\sigma^6}{15 (R-d_{i,j})^6 (R+d_{i,j})^6} \right] , \, d_{ij} < d_{mp}\\
\\
0 , \, d_{ij} > d_{mp}
\end{array}
\right. .
\end{equation}
\end{widetext}
where $R = \frac{a}{2}$ is the particle radius, $A_{mp} = 75.398 \, k_B T$ and $d_{mp}$ is the relative potential cut-off.
Since we model {\it non-sticky} particles, we took $d_{mp}$ in correspondance of the minimum of $U_{mp}$.

$U_{pp}$ is given by:
\begin{equation}\label{eq:Ucc}
U_{pp}(i,j) = \left\{
\begin{array}{l} U_{a, pp}(i,j) + U_{r, pp}(i,j) , \, d_{ij} < d_{pp}\\
\\
0 , \, d_{ij} > d_{pp}
\end{array}
\right.
\end{equation}
where:
\begin{equation}\label{eq:Uacc}
U_{a, pp}(i,j) = - \frac{A_{pp}}{6} \left[ \frac{2 R^2}{d_{i,j}^2 - 4 R^2} + \frac{2 R^2}{d_{i,j}^2} + \ln\left( \frac{d_{i,j}^2-4 R^2}{d_{i,j}^2} \right)\right]
\end{equation}
is the attractive part, and
\begin{widetext}
\begin{equation}\label{eq:Urcc}
U_{r, pp}(i,j) = \frac{A_{pp}}{37800} \frac{\sigma^6}{r} \left[ \frac{d_{i, j}^2 - 14 R d_{i, j} + 54 R^2}{(d_{i, j} - 2R)^7}
                + \frac{d_{i, j}^2 + 14 R d_{i, j} + 54 R^2}{(d_{i, j} + 2R)^7} - 2 \frac{d_{i, j}^2 - 30 R^2}{d_{i, j}^7} \right]
\end{equation}
\end{widetext}
is the repulsive part.
Here, $A_{pp} = 39.478 k_B T$ and $d_{pp}$ is the relative cut-off, again taken at the correspondence of the minimum of $U_{pp}$.

Values of $d_{mp}$ and $d_{pp}$ as functions of particle diameter $a$ are summarized in Table~\ref{tab:SystemsDetails}.
Notice, that for $a = 1.0\sigma$ particle-particle and monomer-particle excluded volume interactions simply reduce
to the same Lennard-Jones function as of monomer-monomer interaction, Eq.~\ref{eq:ljpot}.

\begin{table}
\begin{tabular}{|c|c|c|c|c|c|}
\hline
\multicolumn{2}{|c|}{$a$} & $d_{mp}$ & $d_{pp}$ & $D_{\infty}$ & $\eta_{\infty}$ \\
\hline
$[\sigma]$ & $[\mbox{nm}]$ & $[\sigma]$ & $[\sigma]$ & $\times 10^{-3} [\mu \mbox{m}^2{\mbox{/sec}}]$ & $[\mbox{Pa} \cdot \mbox{s}]$ \\
\hline
$ 1.0$ & $ 30$ &      -- &       -- & $50.0$ & $0.21$\\
$ 2.0$ & $ 60$ & $1.880$ & $ 2.635$ & $28.0$ & $0.25$\\
$ 4.0$ & $120$ & $2.865$ & $ 4.602$ & $10.0$ & $0.42$\\
$ 6.0$ & $180$ & $3.862$ & $ 6.591$ & $ 3.3$ & $0.92$\\
$ 8.0$ & $240$ & $4.861$ & $ 8.585$ & $ 1.3$ & $1.81$\\
$10.0$ & $300$ & $5.860$ & $10.581$ & $ 0.5$ & $3.86$\\
\hline
\end{tabular}
\caption{
\label{tab:SystemsDetails}
Summary of parameters used in this work.
$a$:
diameter of dispersed particles expressed in Lennard-Jones ``$\sigma = 30$ nm'' and ``nm'' units.
$d_{mp}$ and $d_{pp}$:
cut-off distances for monomer-particle (Eq.~\ref{eq:Umc})
and particle-particle (Eq.~\ref{eq:Ucc}) interaction terms, respectively.
For $a = 1.0 \sigma = 30$ nm, monomer-particle and particle-particle excluded volume interactions
reduce to the same functional form as of monomer-monomer interaction, Eq.~\ref{eq:ljpot}.
$D_{\infty} \equiv \lim_{\tau \rightarrow \infty}\frac{\delta x^2 (\tau)}{6 \tau}$ is the particle terminal diffusion coefficient.
Values for different $a$ are obtained as best fits to data reported in Fig.~\ref{fig:monomerMSD}B on the time window $[10^3-10^4]$ seconds.
$\eta_{\infty} \equiv \frac{k_B T}{2 \pi (a+\sigma) D_{\infty}}$ is the corresponding terminal viscosity.
}
\end{table}

\subsection{Simulation protocol II. Molecular Dynamics simulations}\label{sec:MDdetails}

As in our previous studies~\cite{RosaPLOS2008,RosaBJ2010,DiStefanoRosa2013},
polymer/particle dynamics was studied by using fixed-volume Molecular Dynamics simulations at
{\it fixed monomer density} $\rho = N / V = 0.1 / \sigma^3$ with periodic boundary conditions~\cite{PBCnote}.
$V$ is the volume of the region of the box accessible to polymer {\it i.e.} not occupied by the dispersed particles,
hence the total volume of the simulation box is $V_{box} = V + \frac{4 \pi}{3} N_p R^3$.
With this choice:
(1)
polymer density matches the nominal nuclear DNA density of $\approx 0.012 \mbox{ bp} / \mbox{nm}^3$
which was used in our previous studies~\cite{RosaPLOS2008,RosaBJ2010,DiStefanoRosa2013},
and
(2)
the entanglement length $L_e = 1.2 \mbox{ }\mu\mbox{m} = 1.2 \times 10^5 \mbox{ bps}$
and the corresponding tube diameter (mesh size) $d_T = \sqrt{\frac{l_K L_e}{6}} \approx 245$ nm~\cite{RosaPLOS2008,RosaBJ2010,RosaEveraersPRL2014},
of the polymer solution (which are functions of fiber stiffness and density~\cite{uchida})
are not affected by particles insertion.
Interestingly, as reported by a recent study~\cite{KuhnholdPaul_PRE2014} on microrheology of unentangled polymer melts,
simulations at fixed $V$ also guarantee that loss and storage moduli are minimally perturbed by the insertion of dispersed particles.

The system dynamics was integrated by using LAMMPS~\cite{lammps}
with Langevin thermostat in order to keep fixed the temperature of the system to $1.0 k_B T$.
The elementary integration time step is equal to $t_{int} = 0.012 \tau_{MD}$,
where $\tau_{MD}=\sigma(m/\epsilon)^{1/2}$ is the Lennard-Jones time,
$m = M = 1$ are the chosen values~\cite{EqualMassesJustification} for the mass of monomers and particles, respectively.
$\gamma = 0.5 / \tau_{MD}$ is the monomer/particle friction coefficient~\cite{KremerGrestJCP1990}
which takes into account the corresponding interaction with a background implicit solvent.
The total numerical effort corresponds to $2.52 \times 10^7 \tau_{MD}$ per single run,
amounting to $\approx 10^4$ hours on single typical CPU.
The first $1.2 \times 10^6$ MD time steps have been discarded from the analysis of results.

\subsection{Simulation protocol III. Initial configuration}\label{sec:IniConfig}

{\it Construction of model chromosome conformation} --
In spite of the complexity of the chromatin fiber and the nuclear medium,
three-dimensional chromosome conformations are remarkably well described by generic polymer models~\cite{Grosberg_PolSciC_2012,GrosbergKremerRev_RPP2014,RosaZimmer2014}.
In particular, it was suggested~\cite{RosaPLOS2008,VettorelPhysToday2009} that the experimentally observed~\cite{hic} crumpled chromosome
structure can be understood as the consequence of slow equilibration of chromatin fibers due to mutual chain uncrossability during thermal motion.
As a consequence, chromosomes do not behave like equilibrated {\it linear} polymers in solution~\cite{DoiEdwards,RubinsteinColby}.
Instead, they appear rather similar to unlinked and unknotted circular (ring) polymers in entangled solution.
In fact, under these conditions ring polymers are known to spontaneously segregate and
form compact conformations~\cite{VettorelPhysToday2009,GrosbergKremerRev_RPP2014,RosaEveraersPRL2014},
strikingly similar to images of chromosomes in live cells obtained by fluorescence techniques~\cite{CremerReview2001}.

Due to the typical large size of mammalian chromosomes ($\sim 10^8$ basepairs of DNA),
even minimalistic computational models would require the simulation of large polymer chains,
with tens of thousands of beads or so~\cite{RosaPLOS2008,RosaBJ2010,DiStefanoRosa2013}.
For these reasons, in this work we resort to our recent mixed Monte-Carlo/Molecular Dynamics multi-scale algorithm~\cite{RosaEveraersPRL2014},
in order to design a single, equilibrated ring polymer conformation at the nominal polymer density of $\rho = 0.1 / \sigma^3$ (Sec.~\ref{sec:MDdetails}).
The ring is constituted by $N=39154$ monomer particles, which correspond to the average linear size of a mammalian chromosome with $\approx 1.18 \times 10^8$ basepairs.
By construction, the adopted protocol guarantees that the polymer has the nominal local features of the 30nm-chromatin fiber
(stiffness, density and topology conservation) that have been already employed elsewhere~\cite{RosaPLOS2008,RosaBJ2010,DiStefanoRosa2013}.
For the details of the multi-scale protocol, we refer the reader to Ref.~\cite{RosaEveraersPRL2014}.

{\it Insertion of probe particles} --
In order to place $N_p = 100$ spherical particles of increasing radii inside the chromatin solution,
we have proceeded as follows.
First, we have inserted particles of radius $ = 1.0\sigma = 30$ nm at random positions inside the simulation box.
We have carefully removed next unwanted overlaps with chromatin monomers by a short MD run ($\approx 100 \tau_{MD}$)
with the LAMMPS option NVE/LIMIT,
which limits the maximum distance a particle can move in a single time-step, see~\cite{LammpsWebsite}.
At the end of this run, we have gently inflated the simulation box so to reestablish the correct polymer density of $\rho = 0.1/\sigma^3$.
Initial configurations with particles of larger radius are obtained from initial configurations with particles with the immediately smaller radius,
by making use again of the NVE/LIMIT option in order to gently remove possible overlaps, followed again by gentle inflation of the simulation box.

\subsection{Particle-tracking microrheology}\label{sec:ParticleTrackingMicroRheol}

Microrheology~\cite{MasonWeitz1995} employs the diffusive thermal motion of particles dispersed in a medium
in order to derive the complex shear modulus ${\hat G}(\omega) = G'(\omega) + i G''(\omega)$ of the medium~\cite{MasonWeitz1995}.
Following Mason and Weitz~\cite{MasonWeitz1995},
the motion of each dispersed particle is described by a generalised Langevin equation:
\begin{equation}\label{eq:genLangevin}
M \frac{d {\mathbf v}(t)}{d t} = -\int_0^t \gamma(t-\tau) {\mathbf v}(\tau) d\tau + {\mathbf f}(t) ,
\end{equation}
where
$M$ is the mass of the particle,
$\mathbf v$ is its velocity,
${\mathbf f}(t)$ represents the stochastic force acting on the particle consequent from its interaction with the surrounding visco-elastic medium,
and the function $\gamma(t)$ represents the (time-dependent) memory kernel.
Eq.~\ref{eq:genLangevin} is complemented by the fluctuation-dissipation relation:
\begin{equation}\label{eq:fluctdiss}
\langle {\mathbf f}(t) \cdot {\mathbf f}(t') \rangle = 6 \, k_B \, T \, \gamma(t-t') \, .
\end{equation}
By taking the Laplace transform~\cite{LTransformNote} of Eq.~\ref{eq:genLangevin} one gets:
\begin{equation}\label{eq:genLangevinLT1}
M s \, {\tilde {\mathbf v}}(s) - M {\mathbf v}(0) = - {\tilde \gamma}(s) \, {\tilde {\mathbf v}}(s) + {\tilde {\mathbf f}}(s) ,
\end{equation}
or
\begin{equation}\label{eq:genLangevinLT2}
{\tilde {\mathbf v}}(s) = \frac{ M {\mathbf v}(0) + {\tilde {\mathbf f}}(s)}{M s + {\tilde \gamma}(s)} \, .
\end{equation}
Then, by multiplying Eq.~\ref{eq:genLangevinLT2} by ${\mathbf v}(0)$ and taking the thermal average we get:
\begin{equation}\label{eq:genLangevinLT3}
\langle {\tilde {\mathbf v}}(s) \cdot {\mathbf v}(0) \rangle
= \frac{ M \langle {\mathbf v}(0)^2 \rangle + \langle {\tilde {\mathbf f}}(s) \cdot {\mathbf v}(0) \rangle}{M s + {\tilde \gamma}(s)}
= \frac{3 k_B T}{M s + {\tilde \gamma}(s)} \, ,
\end{equation}
where we have used the equipartition relation $M \langle {\mathbf v}(0)^2 \rangle = 3 k_B T$
and the result $\langle {\tilde {\mathbf f}}(s) \cdot {\mathbf v}(0) \rangle = 0$~\cite{KuhnholdPaul_PRE2014}.
The average thermal motion of the dispersed particle is defined through the time mean-square displacement,
$\delta x^2(\tau) \equiv \langle ( {\mathbf x}(t+\tau) - {\mathbf x}(t) )^2 \rangle$,
where ${\mathbf x}(t) = \int_0^t {\mathbf v}(t') \, dt' + {\mathbf x}(0)$ is particle position at time $t$.
Since
$\delta x^2(\tau) = 2 \int_0^\tau (\tau-t) \langle {\mathbf v}(t) \cdot {\mathbf v}(0) \rangle dt$, or
$\delta {\tilde x}^2(s) = \frac{2}{s^2} \langle {\tilde {\mathbf v}}(s) \cdot {\mathbf v}(0) \rangle$,
the Laplace transform of the memory kernel ${\tilde \gamma}(s)$ can be expressed as a function of the Laplace transform of the mean-square displacement, $\delta {\tilde x}^2(s)$:
\begin{equation}\label{eq:memKern}
{\tilde \gamma}(s) = \frac{6 k_B T}{s^2 \, \delta {\tilde x}^2(s)} - M s \, .
\end{equation}
By assuming~\cite{MasonWeitz1995} that ${\tilde \gamma}(s)$
is proportional to the bulk frequency-dependent viscosity of the fluid, ${\tilde \eta}(s)$, we get finally:
\begin{equation}\label{eq:MicrMemFunct}
{\tilde \eta}(s) = \frac{{\tilde \gamma}(s)}{\nu \pi a} ,
\end{equation}
as in the case of a standard viscous fluid.
The parameter $\nu$ depends on the boundary condition at the particle surface~\cite{OuldKaddourLevesque_PRE63_2000}:
for sticky boundary condition $\nu=3$ and for slip boundary condition $\nu=2$.
The Laplace-transform of the {\it shear modulus}
${\tilde G}(s) = s \, {\tilde \eta}(s)$ is given by:
\begin{equation}\label{eq:ShearModulus}
{\tilde G}(s)
=       \frac{s}{\nu \pi a} \left[ \frac{6 k_B \, T}{s^2 \, \delta {\tilde x}^2(s)} - M \, s \right]
\approx \frac{6 \, k_B \, T}{\nu \pi a \, s \, \delta {\tilde x}^2(s)} \, ,
\end{equation}
where the last expression is obtained by neglecting the inertia term~\cite{MasonWeitz1995}.
Finally, the complex shear modulus ${\hat G}(\omega)$ as a function of frequency $\omega$
is obtained from ${\tilde G}(s)$ by analytical continuation upon substitution of ``$s$'' with ``$i \omega$'':
${\hat G}(\omega) =  -i \frac{6 \, k_B \, T}{\nu \pi a \, \omega \, \delta {\tilde x}^2(s = i\omega)}$.
Its real ($G'(\omega)$) and imaginary ($G''(\omega)$) parts correspond to the so-called storage and loss moduli
and are a measure of the elastic and viscous properties of the solution~\cite{RubinsteinColby}, respectively.

\section{Results}\label{sec:Results}
\begin{figure}
\includegraphics[width=3.3in]{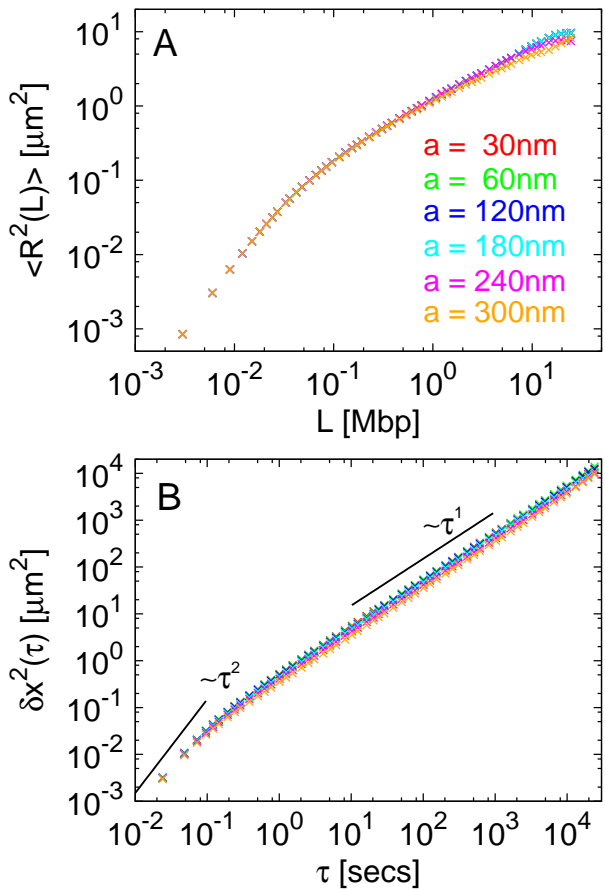}
\caption{
\label{fig:SimulsTest}
(A)
Average-square internal distances, $\langle R^2(L) \rangle$, between pairs of monomers at genomic separation, $L$,
along the ring.
$L$ is taken up to $1/4$ of the entire ring contour length ($\approx 118$ Mbp).
(B)
Time mean-square displacement $\delta x^2(\tau)$ of dispersed particles in the absence of the polymer.
Color code is as in panel A.
}
\end{figure}

Before proceeding to analyse our results on the diffusion of particles dispersed in the chromatin solution,
we validated our system setup whether
(1) the chromosome conformation is not perturbed by the insertion of particles,
and
(2) particles diffusion is exclusively influenced by particle-chromatin interaction and not by particle-particle interactions.
In order to test (1),
as a measure of chromosome conformation we considered the average-square internal distances $\langle R^2(L) \rangle$~\cite{RosaPLOS2008}
between pairs of chromatin beads at genomic distance, $L$.
Fig.~\ref{fig:SimulsTest}A shows plots of $\langle R^2(L) \rangle$ for all sizes of dispersed particles.
The almost perfect match between different curves shows that, on average, the polymer maintains the same spatial conformation.
In order to test (2),
we removed the chromosome and measured particle diffusion then.
Fig.~\ref{fig:SimulsTest}B shows that particles diffuse normally with diffusion coefficients barely depending on particle size.
Hence, even if particle-particle collisions were not completely excluded from our system, their effect is small,
in particular it is much smaller than the effect due to collisions between particles and chromatin monomers (see Fig.~\ref{fig:monomerMSD} below).
Of course, the effect of particle-particle collisions can be reduced even further by considering a larger polymer system at the same number of diffusing particles, $N_p$.

\begin{figure*}
\includegraphics[width=6.6in]{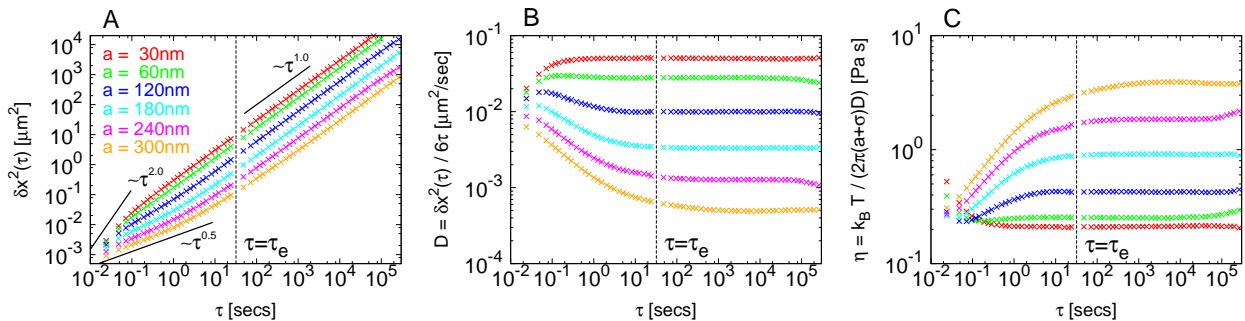}
\caption{
\label{fig:monomerMSD}
(A)
Time mean-square displacement, $\delta x^2(\tau)$, of particles of diameter $a$ dispersed in the polymer (chromatin) solution.
At short times and for large particle sizes, we observe an anomalous behavior
$\delta x^2(\tau) \approx \tau^{\alpha}$, with $\alpha$ slowly approaching $0.5$ as predicted by Cai {\it et al.}~\cite{CaiPanuykovRubinstein2011}.
For comparison, vertical dashed lines mark the position of the nominal entanglement time $\tau_e \approx 32$ seconds~\cite{RosaPLOS2008}
of chromatin solution.
(B)
Corresponding diffusion coefficient, $D(\tau) = \delta x^2(\tau) / 6 \tau$.
At lag-times longer than $\tau_e$, diffusion becomes normal and $D$ takes a constant value.
(C)
Corresponding viscosities, $\eta(\tau) = \frac{k_B T}{2 \pi \, (a+\sigma) \, D(\tau)}$.
}
\end{figure*}

After validation of our model, we proceeded to analyse the dynamics of dispersed particles
in the presence of the polymer and as a function of particle diameter, $a$.
We considered first the time mean-square displacement,
$\delta x^2(\tau)$, of the particles,
and the corresponding instantaneous diffusion coefficient and viscosity respectively defined
as $D(\tau) = \frac{\delta x^2(\tau)}{6 \, \tau}$ and $\eta(\tau) = \frac{k_B T}{2 \pi \, (a+\sigma) \, D(\tau)}$,
where ``$a+\sigma$'' is the cross-diameter of the dispersed particle and
the slip boundary condition applies~\cite{OuldKaddourLevesque_PRE63_2000}.

Table~\ref{tab:SystemsDetails} and Fig.~\ref{fig:monomerMSD} summarize our results.
We found that, after a short ballistic time regime, small particles with $a=30$ nm and $a=60$ nm
diffuse normally, $\delta x^2(t) = 6 D_{\infty} t$,
with terminal diffusion coefficients
$D_{\infty} \equiv \lim_{\tau \rightarrow \infty} \frac{\delta x^2(\tau)}{6 \tau} \approx 5 \times 10^{-2} \, {\mu \mbox{m}}^2 / \mbox{s}$
and $D_{\infty} \approx 3 \times 10^{-2} \, {\mu \mbox{m}}^2 / \mbox{s}$, respectively,
corresponding to terminal viscosities $\eta_{\infty} \equiv \frac{k_B T}{2 \pi \, (a+\sigma) D_{\infty}} \approx 0.21$ Pa$\cdot$s
and $\eta_{\infty} \approx 0.25$ Pa$\cdot$s.
As particle size was increased from $a=120$ to $a=300$ nm,
we observed:
(1) the appearance of a small-time anomalous regime $\delta x^2(\tau) \sim \tau^{\alpha}$ with $\alpha$ slowly approaching $0.5$~\cite{CaiPanuykovRubinstein2011} for large $a$,
(2) a dramatic drop in the terminal diffusion coefficient down to $\approx 5 \times 10^{-4} \, {\mu \mbox{m}}^2 / \mbox{s}$, and
(3) an increase of the corresponding viscosity with particle size up to $\approx 4$ Pa$\cdot$s.

\begin{figure}
\includegraphics[width=3.5in]{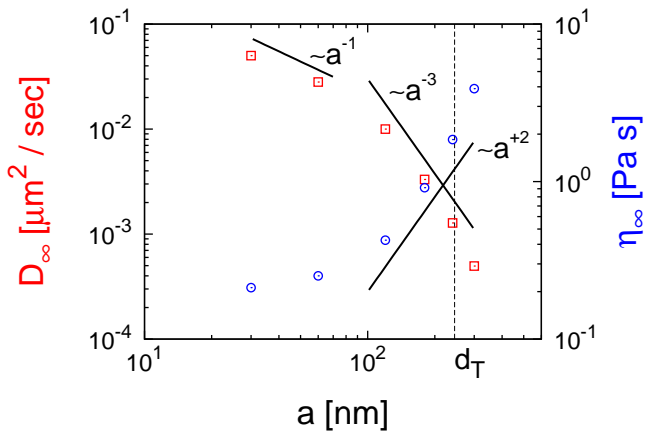}
\caption{
\label{fig:DiffCoeffViscosity_VS_Diam}
Long-time limit of particle diffusion coefficient, $D_{\infty}$ ($\square$), and particle viscosity, $\eta_{\infty}$ ($\circ$),
as a function of particle diameter, $a$.
Power-laws correspond to theoretical predictions by Cai {\it et al.}~\cite{CaiPanuykovRubinstein2011}.
In particular, deviation from the intermediate power-law behavior $\sim a^{-3}$ for $D_{\infty}$ and $\sim a^2$ for ${\eta}_{\infty}$
arise when particle size becomes comparable to the tube diameter of chromatin solution, $d_T \approx 245$ nm~\cite{RosaPLOS2008}.
}
\end{figure}

We interpreted our results
at the light of the scaling argument discussed by Cai {\it et al.}~\cite{CaiPanuykovRubinstein2011}.
As in the case of any general polymer solution, our model chromatin mesh can be characterized by two fundamental quantities~\cite{DoiEdwards,RubinsteinColby,RosaPLOS2008}:
(1) the fiber stiffness, measured in terms of the Kuhn length $l_K = 300 \mbox{ nm} = 3 \times 10^4$ basepairs, and
(2) the entanglement length, $L_e = 1.2 \, \mu \mbox{m} = 1.2 \times 10^5$ basepairs,
which is a function of chromatin stiffness and density~\cite{uchida,RosaPLOS2008}
and represents the characteristic chain contour length value above which polymers start to entangle.
Kinetic properties of the chromatin solution are affected by entanglements on
length scales larger than the so-called tube diameter (or mesh size) of the solution, $d_T = \sqrt{\frac{l_K L_e}{6}} \approx 245$ nm,
and on time scales larger than the entanglement time, $\tau_e \approx 32$ seconds~\cite{RosaPLOS2008}.
According to Cai {\it et al.}~\cite{CaiPanuykovRubinstein2011},
since particle size is at most only slightly larger than $d_T$ (see Table~\ref{tab:SystemsDetails}),
entanglements are not expected to affect significantly particle diffusion.
Under these conditions, particle dynamics is coupled to the ``Rouse-like'' relaxation modes of
chromatin segments with contour length shorter than $L_e$,
namely made of $n_K(\tau) \sim (\tau / \tau_K)^{1/2}$ Kuhn segments and having spatial size $\sim l_K \, n_K(\tau)^{1/2} \sim l_K \, (\tau / \tau_K)^{1/4}$.
The corresponding chromatin viscosity is given~\cite{RubinsteinColby} then by
$\eta(\tau) \sim \eta_K \, n_K(\tau) \sim \eta_K (\tau / \tau_K)^{1/2}$ where
$\eta_K$ and $\tau_K$ are the viscosity and relaxation time of a Kuhn segment, respectively.
The mean-square displacement of the particle is then given by
$\delta x^2(\tau) \sim \frac{k_B T}{a \, \eta(\tau)} \, \tau \sim \frac{k_B T}{a \, \eta_K} \, (\tau \, \tau_K)^{1/2}$,
up to time-scale $\tau_r$
where polymer sections become comparable to particle size
$l_K \, (\tau_r / \tau_K)^{1/4} \sim a$ or
$\tau_r \sim \tau_K (a/l_K)^4$.
On longer time-scales, particle displacement is normal,
$\delta x^2(\tau) \sim \frac{k_B T}{a \, \eta(\tau_r)} \, \tau$,
with terminal diffusion coefficient $D_{\infty} \sim \frac{k_B T}{a \, \eta(\tau_r)} \sim 1 / a^3$ and viscosity $\eta_{\infty} \sim \frac{1}{a \, D_{\infty}} \sim a^2$.
Fig.~\ref{fig:DiffCoeffViscosity_VS_Diam} summarizes our results for $D_{\infty}$ and $\eta_{\infty}$
showing good agreement with theoretical predictions.
Nonetheless, we report some deviation from the predicted behavior when the particle size reaches the nominal tube diameter $d_T \approx 245$ nm
(rightmost symbols in Fig.~\ref{fig:DiffCoeffViscosity_VS_Diam}).

\begin{figure*}
\includegraphics[width=6.6in]{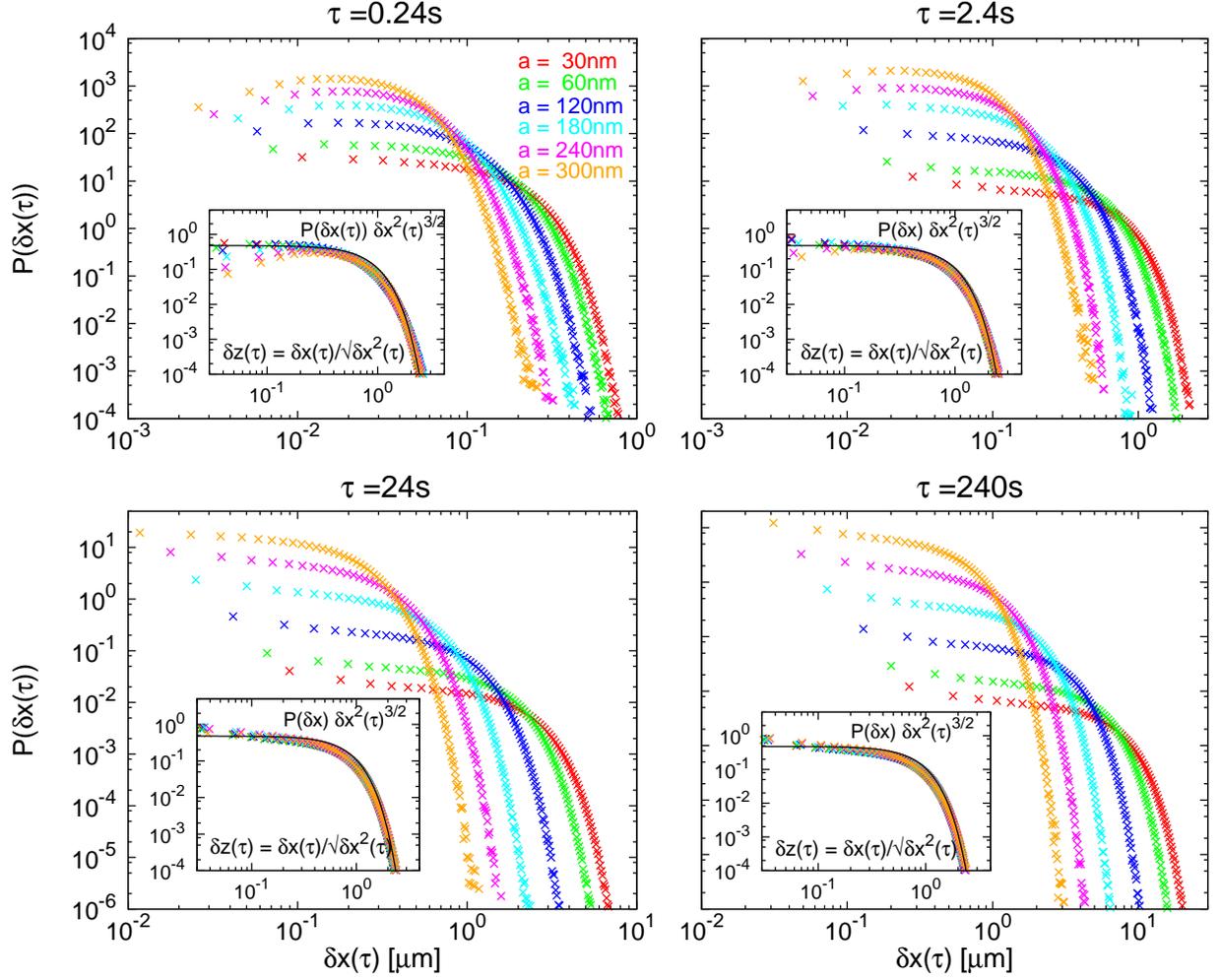}
\caption{
\label{fig:jumpPDF}
Distribution function, $P({\delta x(\tau)})$, of particle displacements, $\delta x(\tau) = |{\mathbf x}(t+\tau) - {\mathbf x}(t)|$,
for different values of lag-times $\tau$ (see captions).
Insets: data rescaled according to
$\delta z(\tau) = \delta x(\tau) / \sqrt{\delta x^2(\tau)}$ and
$P({\delta z(\tau)}) = P({\delta x(\tau)}) \, \delta x^2(\tau)^{3/2}$.
At long lag-times, rescaled data are described by a universal Gaussian function
$P(\delta z(\tau)) = \left( \frac{3}{2 \pi} \right)^{3/2} \exp \left( - \frac{3}{2} \delta z(\tau)^2 \right)$ (black solid lines), as expected.
At short lag-times $\tau$ and large particle sizes, $P(\delta z(\tau))$ deviates significantly from Gaussian behavior.
}
\end{figure*}

\begin{figure}
\includegraphics[width=3.25in]{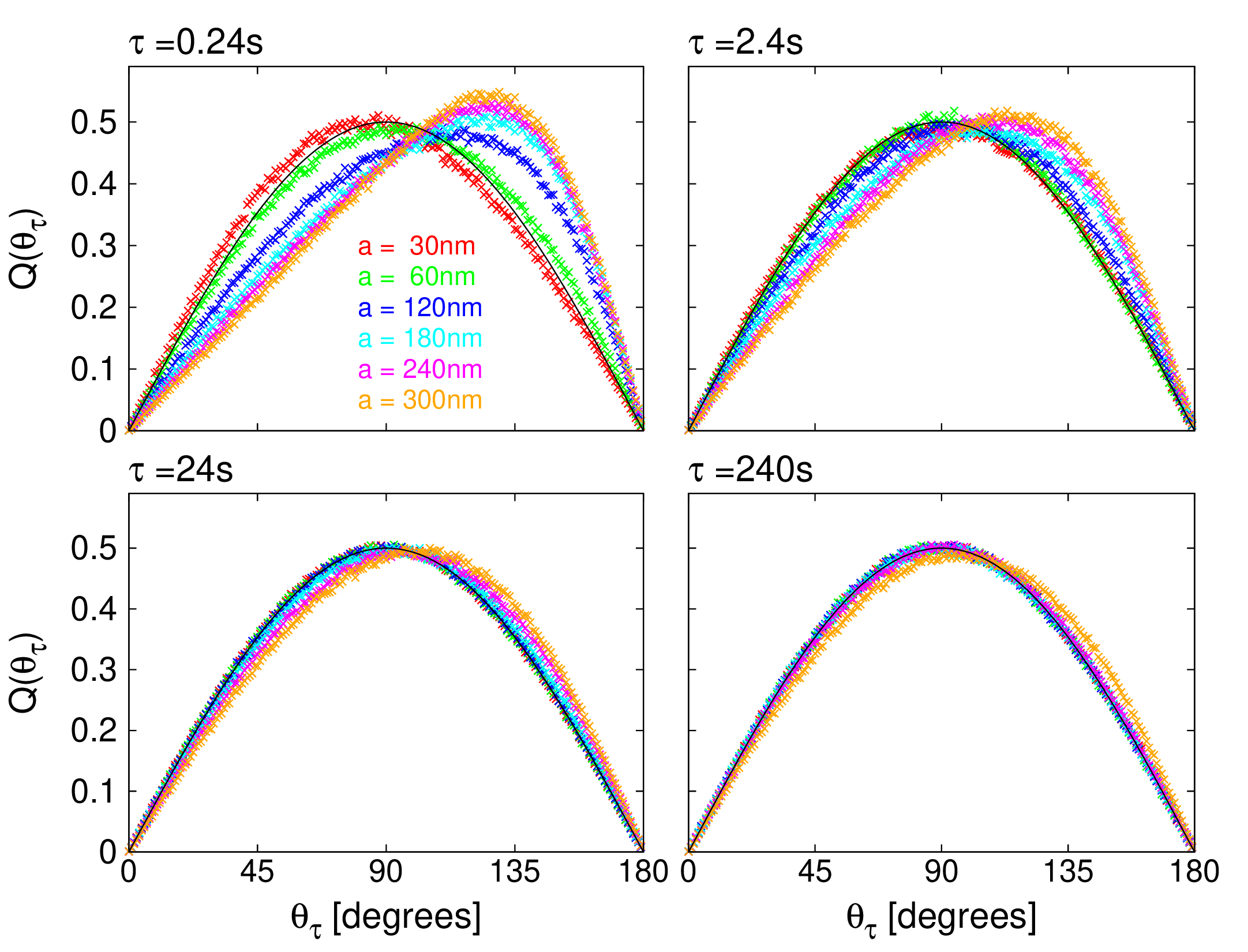}
\caption{
\label{fig:AngleJumpsPDF}
Distribution function, $Q({\theta}_{\tau})$, of angles $\theta_{\tau}$ between $\tau$-lagged particle vector displacements
${\mathbf x}(t+\tau) - {\mathbf x}(t)$ and ${\mathbf x}(t+2\tau) - {\mathbf x}(t+\tau)$ taken consecutively along the trajectory.
Values of $\tau$ are as in Fig.~\ref{fig:jumpPDF}.
At small $\tau$ and large particles, significant deviations from the random distribution $\sin(\theta_{\tau})/2$ (black solid lines) appear.
}
\end{figure}

In spite of the reported evidence that particle dynamics appears to be dominated by the relaxational Rouse modes of chromatin linear sections $< L_e$,
we found that entanglements play nonetheless a (although quite more subtle) role:
in fact, they lead to the formation of effective ``domains'' which cage the particles.
In order to show that, we computed the probability distribution functions, $P(\delta x(\tau))$,
of particle displacements $\delta x(\tau) \equiv {\mathbf x}(t+\tau) - {\mathbf x}(t)$ at lag-times $\tau$.
To fix the ideas, we chose lag-times $\tau = 0.24, 2.4, 24, 240$ s for all particle sizes, see Fig.~\ref{fig:jumpPDF}.
Interesting features emerge upon calculation of the corresponding scaling plots (insets) obtained by substitutions
$\delta x(\tau) \rightarrow \delta x(\tau) / \sqrt{\delta x^2(\tau)}$ and
$P({\delta x(\tau)}) \rightarrow P({\delta x(\tau)}) \, \delta x^2(\tau)^{3/2}$.
In particular, at long lag-times and for all particle sizes, the distribution is described by a simple Gaussian,
$P(\delta x(\tau)) = \left(\frac{3}{2 \pi \, \delta x^2(\tau)}\right)^{3/2} \exp \left( -\frac{3 \, (\delta x(\tau))^2}{2 \, \delta x^2(\tau)} \right)$ (black lines).
At short lag-times, the Gaussian distribution holds for small particles only.
In fact, at large particle sizes and small $\delta x(\tau)$,
$P(\delta x(\tau))$ shows significant deviations from the Gaussian behavior
which can be understood in terms of partial trapping due to the emerging topological constraints~\cite{KalathiGrest_PRL2014}.
This conclusion is further supported (see Fig.~\ref{fig:AngleJumpsPDF}) by the behaviour of the distribution function $Q(\theta_{\tau})$ of angles
$\theta_{\tau} \equiv \cos^{-1} \left( \frac{ \left( {\mathbf x}(t+\tau) - {\mathbf x}(t) \right) \cdot \left( {\mathbf x}(t+2\tau) - {\mathbf x}(t+\tau) \right) } { \left| {\mathbf x}(t+\tau) - {\mathbf x}(t) \right| \, \left| {\mathbf x}(t+2\tau) - {\mathbf x}(t+\tau) \right| } \right)$
between $\tau$-lagged particle vector displacements taken consecutively along the trajectory.
At $\tau=0.24$ s, $Q(\theta_{\tau})$ for small particles diameters of $30$ and $60$ nm match the random distribution
$Q(\theta_{\tau}) = \sin(\theta_{\tau})/2$ (black solid lines in Fig.~\ref{fig:AngleJumpsPDF}).
For larger particle sizes, $Q(\theta_{\tau})$ is constantly shifted towards higher values of $\theta_{\tau}$,
which is compatible with the picture where particles revert their motion frequently as the consequence of trapping inside chromatin domains.
Finally, at large $\tau$ where particle motion is diffusive for all particle sizes, $Q(\theta_{\tau})$ becomes compatible again with the random distribution
(see corresponding panels in Fig.~\ref{fig:AngleJumpsPDF}), as expected.

\begin{figure*}
\includegraphics[width=6.7in]{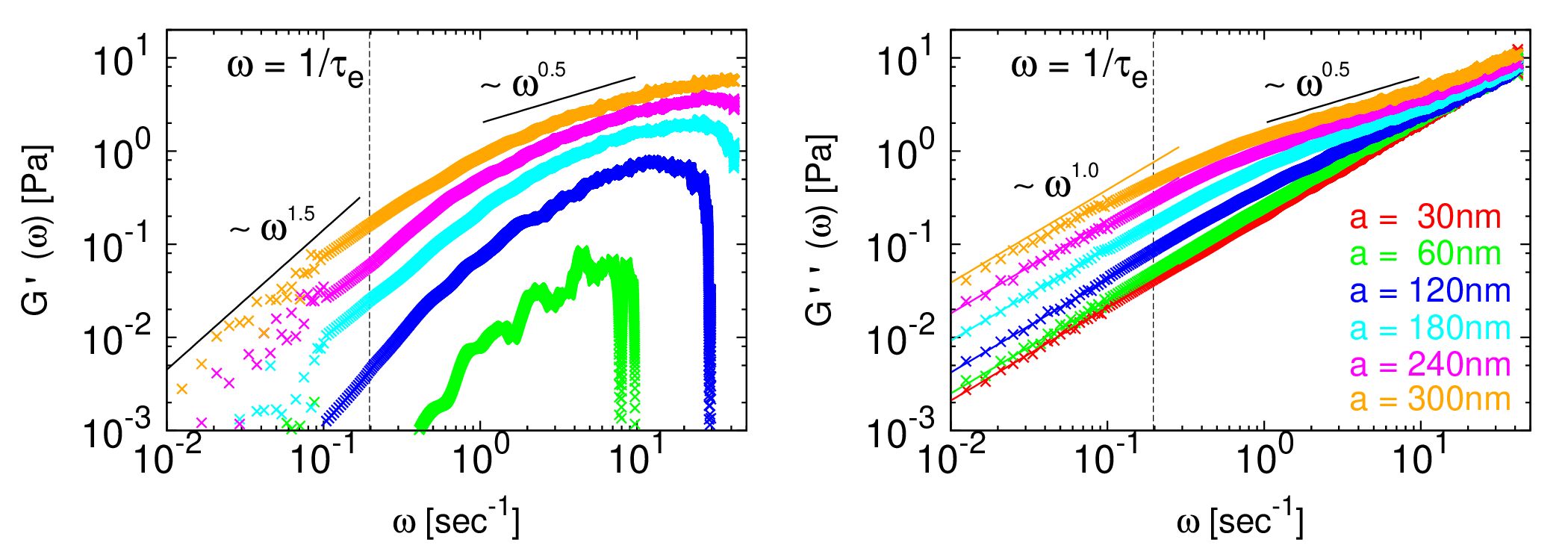}
\caption{
\label{fig:Gomega}
Storage, $G'(\omega)$, and loss, $G''(\omega)$, moduli as a function of frequency $\omega$.
For small frequencies, $G''(\omega) \approx \eta_{\infty} \omega$ (solid lines) with $G''(\omega) \gg G'(\omega) \sim \omega^{1.5}$,
{\it i.e.} the medium responds as a standard viscous fluid, see Eq.~\ref{eq:GoneGtwoOmega0} with $\alpha=0.5$.
For large frequencies and large length-scales,
the elastic modulus becomes comparable to the loss modulus $G'(\omega) \approx G''(\omega) \approx \omega^{1/2}$
as expected, see Eq.~\ref{eq:GoneGtwoOmegaInf} with $\alpha=0.5$.
Numerical data for $G'$ and $G''$ have been obtained from numerical calculation of $\delta {\tilde x}^2(s=i\omega)$,
Eq.~\ref{eq:Tassieri}, followed by smoothing~\cite{EvansTassieri2009} for better visualization.
Vertical dashed lines mark the position of the frequency $\tau_e^{-1}$ equal to the inverse of the entanglement time $\tau_e \approx 32$ seconds, see Fig.~\ref{fig:monomerMSD} for comparison.
}
\end{figure*}

To complete our analysis,
we calculated the storage ($G'(\omega)$) and loss ($G''(\omega)$) modulus of the chromatin solution from, respectively, the real and imaginary part of the complex shear modulus
${\hat G}(\omega) = -i \frac{3 \, k_B \, T}{\pi (a+\sigma) \, \omega \, \delta {\tilde x}^2(s = i \omega)}$, see Sec.~\ref{sec:ParticleTrackingMicroRheol}.
For the generic case where particle motion is subdiffusive at short times and diffusive at large times,
the mean-square displacement can be phenomenologically described by
$\delta x^2(\tau) = 6 D_{\alpha} \, \tau^{\alpha} + 6 D_{\infty} \, \tau $ where $D_{\alpha}$
is the (generalized) diffusion coefficient associated to the anomalous time regime with exponent $0 < \alpha < 1$.
In this case, $\delta {\tilde x}^2(s) = 6 D_{\alpha} \, \Gamma(\alpha + 1) \, s^{-(\alpha+1)} + 6 D_{\infty} \, s^{-2}$.
After some algebra, $G'(\omega)$ and $G''(\omega)$ are given by the following expressions:
\begin{widetext}
\begin{eqnarray}\label{eq:GoneGtwo}
G'(\omega)  & = & \frac{3 \, k_B \, T}{\pi (a+\sigma)} \, \frac{6 D_{\alpha} \, \Gamma(\alpha + 1) \cos(\pi \alpha / 2) \, \omega^{-\alpha}}{\left( 6 D_{\alpha} \, \Gamma(\alpha + 1) \cos(\pi \alpha / 2) \, \omega^{-\alpha} \right)^2 + \left( 6 D_{\alpha} \, \Gamma(\alpha + 1) \sin(\pi \alpha / 2) \, \omega^{-\alpha} + 6 D_{\infty} \, \omega^{-1} \right)^2}\nonumber\\
G''(\omega) & = & \frac{3 \, k_B \, T}{\pi (a+\sigma)} \, \frac{6 D_{\alpha} \, \Gamma(\alpha + 1) \sin(\pi \alpha / 2) \, \omega^{-\alpha} + 6 D_{\infty} \, \omega^{-1}}{\left( 6 D_{\alpha} \, \Gamma(\alpha + 1) \cos(\pi \alpha / 2) \, \omega^{-\alpha} \right)^2 + \left( 6 D_{\alpha} \, \Gamma(\alpha + 1) \sin(\pi \alpha / 2) \, \omega^{-\alpha} + 6 D_{\infty} \, \omega^{-1} \right)^2} \, . \nonumber\\
\end{eqnarray}
\end{widetext}
In the limit
$\omega \rightarrow 0$
they simplify to:
\begin{eqnarray}\label{eq:GoneGtwoOmega0}
G'(\omega)  & = & \frac{k_B \, T}{2 \pi (a+\sigma) \, D_{\infty}} \, \frac{D_{\alpha} \, \Gamma(\alpha + 1)}{D_{\infty}} \cos(\pi \alpha / 2) \, \omega^{2-\alpha} \nonumber\\
G''(\omega) & = & \frac{k_B \, T}{2 \pi (a+\sigma) \, D_{\infty}} \, \omega \, .
\end{eqnarray}
Notice, that in this limit $G'(\omega)$ is always $< G''(\omega)$ and that the coefficient of $G''(\omega)$ is just the terminal viscosity, $\eta_{\infty}$.
In the opposite limit
$\omega \rightarrow \infty$
we find instead:
\begin{eqnarray}\label{eq:GoneGtwoOmegaInf}
G'(\omega)  & = & \frac{k_B \, T}{2 \pi (a+\sigma) \, D_{\alpha}} \, \frac{\cos(\pi \alpha / 2)}{\Gamma(\alpha+1)} \, \omega^{\alpha} \nonumber\\
G''(\omega) & = & \frac{k_B \, T}{2 \pi (a+\sigma) \, D_{\alpha}} \, \frac{\sin(\pi \alpha / 2)}{\Gamma(\alpha+1)} \, \omega^{\alpha} \, .
\end{eqnarray}
In this limit, $G'(\omega)$ and $G''(\omega)$ have the same power-law behavior,
with $G'(\omega)$ smaller (resp., larger) than $G''(\omega)$ for $1/2 < \alpha < 1$ (resp., $0 < \alpha < 1/2$).
Our results for the mean-square displacement $\delta x^2(\tau) \sim \tau^{\alpha}$ with $\alpha \gtrsim 0.5$ (Fig.~\ref{fig:monomerMSD}A)
then predict that $G'(\omega) < G''(\omega)$ on the entire frequency range considered here.

In order to derive the complex shear modulus ${\hat G}(\omega)$ of the chromatin solution,
we resorted to the numerical method developed by Evans and coworkers~\cite{EvansTassieri2009,EvansTassieri2010,EvansTassieri2012}.
The method allows a straightforward evaluation of the Laplace transform $\delta {\tilde x}^2(s=i\omega)$ of $\delta x^2(\tau)$
through the formula:
\begin{widetext}
\begin{equation}\label{eq:Tassieri}
-\omega^{2} {\delta {\tilde x}^2}(s=i\omega)
= i \omega \, \delta x^2(0) +
\left( 1 - e^{-i \omega \tau_{1}} \right) \frac{ \left( \delta x^2(\tau_1) - \delta x^2(0) \right) }{\tau_1} +
6 D_{\infty} \, e^{-i \omega \tau_J} + \sum_{j=2}^J \left( \frac{ \delta x^2(\tau_j) - \delta x^2(\tau_{j-1}) } { \tau_j - \tau_{j-1} } \right) \left( e^{-i \omega \tau_{j-1}} - e^{-i \omega \tau_j} \right) \, ,
\end{equation}
\end{widetext}
where
$\left( (\tau=0, \delta x^2(\tau=0)), (\tau=\tau_1, \delta x^2(\tau=\tau_1)), ..., \right.$ $\left. (\tau=\tau_J, \delta x^2(\tau=\tau_J)) \right)$ is the time-series for $\delta x^2(\tau)$.

Results are summarized in Fig.~\ref{fig:Gomega}.
For small frequencies, we confirm the predicted results $G'(\omega) \sim \omega^{1.5} \ll G''(\omega) \approx \eta_{\infty} \omega$,
{\it i.e.} the polymer solution behaves as a typical viscous medium.
For large frequencies, deviation from this behavior is particularly evident in the case of large particles
where $G'(\omega) \lesssim G''(\omega) \sim \omega^{1/2}$,
and the polymer solution behaves like a ``power-law'' liquid.
Representative values for $G'(\omega)$ and $G''(\omega)$ at $\omega = 0.1$, $1$ and $10$ Hz are reported in Table~\ref{tab:GoneGtwoValues}.

\begin{table*}
\begin{tabular}{|c|c|c|c|c|c|c|}
\hline
& \multicolumn{2}{c}{$0.1$ Hz} & \multicolumn{2}{|c|}{$1$ Hz} & \multicolumn{2}{c|}{$10$ Hz}\\
\hline
$a \, [\mbox{nm}]$ & $G' \, [\mbox{Pa}]$ & $G'' \, [\mbox{Pa}]$ & $G' \, [\mbox{Pa}]$ & $G'' \, [\mbox{Pa}]$ & $G' \, [\mbox{Pa}]$ & $G'' \, [\mbox{Pa}]$\\
\hline
$ 30$ & $-$      & $0.0214$ & $-$      & $0.2045$ & $-$      & $2.0651$\\
$ 60$ & $0.0004$ & $0.0238$ & $0.0079$ & $0.2532$ & $-$      & $2.2061$\\
$120$ & $-$      & $0.0423$ & $0.0712$ & $0.3948$ & $0.7009$ & $2.3778$\\
$180$ & $0.0086$ & $0.0882$ & $0.2260$ & $0.6865$ & $1.6126$ & $2.8336$\\
$240$ & $0.0349$ & $0.1469$ & $0.4959$ & $1.0163$ & $2.6690$ & $3.5608$\\
$300$ & $0.0747$ & $0.2972$ & $0.8674$ & $1.4476$ & $3.7901$ & $4.5105$\\
\hline
\end{tabular}
\caption{
\label{tab:GoneGtwoValues}
Representative values of $G'(\omega)$ and $G''(\omega)$ taken at frequencies $\omega = 0.1$, $1$ and $10$ Hz.
Data which are too noisy have not been reported.
}
\end{table*}

\section{Discussion and conclusions}\label{sec:DiscConcls}

The nucleus of eukaryotic cells is a highly crowded medium dominated by the presence of chromatin fibers which form an intricated polymer network.
In this network, several protein complexes diffuse while targeting specific genome sequencing~\cite{alberts}.
In order to understand the mechanisms of macromolecular diffusion inside the nucleus,
the tracking of artificially-designed injected micron-sized particles (microrheology) has been recently introduced~\cite{TsengWirtz2004,WirtzReview}.

In this work, we investigated the diffusion of $100$ tracer particles inside a polymer environment modelling
the structure of interphase chromosomes in mammals.
In particular, we considered particle sizes ranging from $30$ to $300$ nm in order to explore the dynamic response on polymer length-scales from the nominal chromatin diameter
($30$ nm) up to just slight above the so-called mesh (entanglement) size of the chromatin solution, $d_T \approx 245$ nm~\cite{RosaPLOS2008}.

In qualitative agreement with other computational studies~\cite{KalathiGrest_PRL2014,KuhnholdPaul_PRE2014} on the generic behaviour of micro-tracers in model polymer melts,
we found that small particles undergo normal diffusion at all times, while intermediate-size particle sub-diffuse at short times and diffuse normally later on, see Fig.~\ref{fig:monomerMSD}.
In particular, terminal diffusivities can be understood in terms of the scaling theory proposed recently by Cai {\it et al.}~\cite{CaiPanuykovRubinstein2011},
see Fig.~\ref{fig:DiffCoeffViscosity_VS_Diam}.
Of remarkable interest is the dynamic behaviour at times shorter than the entanglement time, $\tau_e$, in particular for big particles.
In fact, the sub-diffusive behaviour appears here to be associated to temporary trapping on scales compatible with the emerging of topological constraints in the underlying chromatin solution, see Fig.~\ref{fig:jumpPDF} and Fig.~\ref{fig:AngleJumpsPDF}.
Interestingly, two independent experimental studies on murine fibroblasts~\cite{TsengWirtz2004} and human HeLa cells~\cite{Shivashankar_PlosOne2012} employing micro-tracers of, respectively, $0.1$ and $1.0$ micron of diameter found particle trapping in nuclear domains of size $\approx 290$ nm and $\approx 250$ nm respectively,
which is in good quantitative agreement with the nominal mesh size of our chromatin solution.
We stress nonetheless, that here we {\it reinterpret} the experimentally observed caging as simply being the consequence of the formation of entanglements in the chromatin fiber, {\it i.e.} as a genuine polymer effect~\cite{DoiEdwards,RubinsteinColby}.

We compare then our predictions for the storage, $G'(\omega)$, and loss, $G''(\omega)$, moduli to corresponding results for fibroblasts and HeLa cells.
For micro-tracers of $0.1 \mu$m-diameter in fibroblasts,
Tseng {\it et al.}~\cite{TsengWirtz2004} found that the nuclear medium responds elastically in the range $[1-10]$ Hz with a plateau modulus
$G'(\omega) \approx 10$ Pa and $G''(\omega)$ in the range $[3-10]$ Pa.
This contrasts with our findings in several respects: our predicted environment is in general much softer (see Table~\ref{tab:GoneGtwoValues}),
and we do not observe any plateau, for our simulated medium results more liquid-like than solid-like.
On the other hand, the work by Hameed {\it et al.}~\cite{Shivashankar_PlosOne2012} on HeLa cells explored by $1 \mu$m-sized particles suggests a much softer
nuclear environment with $G'(\omega) \approx 0.1 \mbox{ Pa} > G''(\omega) \approx 0.05 \mbox{ Pa}$ at $\omega = 1$ Hz.
While still suggesting a nucleus which is more solid- than liquid-like, these results are {\it quantitatively} closer to ours,
although they were obtained with a quite larger tracer bead than the ones used in our simulations.

Quantitative differences between these two experiments can be due to the different cell lines used,
while differences between experiments and theory might be due to the simplicity of the polymer model.
In particular, we would like to stress two important points which were neglected in our study.

First,
a conspicuous number of experimental observations~\cite{CookReview2010,hic,DixonNature2012} demonstrated that
chromatin loci far along the sequence frequently interact with each other because of the presence of specific protein bridges.
From a polymer physics perspective, this creates effective chromatin-chromatin cross-links.
It was suggested~\cite{KalathiGrest_PRL2014} that permanent cross-links in polymer solutions and melts might alter significantly the diffusive behavior
of micro-particles when their size becomes comparable to the polymer mesh size.
Therefore, as a possible avenue for further investigations, it would be interesting to clarify to which extent cross-links added to the system would alter the viscoelastic behaviour reported here for non cross-linked chromatin fibers.

Second,
recent experimental studies~\cite{Shivashankar_PlosOne2012,spakowitzPNAS2012} demonstrated that chromosomal activity and chromosomal loci dynamics
is the result of a subtle interplay between {\it passive} thermal diffusion
and {\it active}, ATP-dependent motion triggered by chromatin remodeling and transcription complexes.
The mentioned work by Hameed {\it et al.}~\cite{Shivashankar_PlosOne2012} showed
that the persistent, caged behavior of the micro-tracers can be altered by imposing an external force on the tracers above a certain threshold
which stimulates frequent jumps between the cages.
Remarkably, these jumps become almost suppressed after ATP-depletion.
This observation seems then to point to the important role played by active mechanisms during micro-tracers dynamics.
Later on in the paper, these mechanisms were ascribed to dynamic remodeling of the chromatin fiber~\cite{Shivashankar_PlosOne2012}.
Consistent with that, Weber {\it et al.}~\cite{spakowitzPNAS2012} showed that diffusion of chromosomal loci in bacteria and yeast is also ATP-dependent.
Taken together, these results suggest that a picture where chromatin fibers are modeled as just as passive polymer filaments is necessarily an approximation.
To move beyond this approximation, recently Ganai {\it et al.}~\cite{ganai2014chromosomeNAR} proposed a novel computational approach
where chromosomes were modeled as chains of beads which were let evolving by a Langevin equation with a non-uniform, monomer-dependent temperature
linking -- at a phenomenological level -- monomer gene content to ``out-of-equilibrium'' activity:
gene-rich monomers are ``hot/active'' while gene-poor monomers are ``cold/passive''.
Interestingly, the work comes to the conclusion that quantitative understanding of the observed chromosomal arrangement inside the nucleus by purely passive mechanisms is incomplete, namely active mechanisms are also needed.
For all these reasons, it would be also interesting to explore in the near future to which extent the viscoelastic properties of active-driven chromatin fibers
deviate from the theoretical predictions presented in this work.

{\it Acknowledgements} --
AR acknowledges
discussions with C. Micheletti and A.-M. Florescu,
and grant PRIN 2010HXAW77 (Ministry of Education, Italy).

%\bibliography{biblio}% Produces the bibliography via BibTeX.

\begin{thebibliography}{41}
\bibitem [{Equ()}]{EqualMassesJustification}%
  \BibitemOpen
  \href@noop {} {}\bibinfo {note} {In this work the monomer mass ($m$) and the
  particle mass ($M$) were tuned to the same value $=1$. This choice was
  motivated by reasons of pure convenience, as in the subsequent analysis the
  physically irrelevant (and, mass-dependent) ballistic regime for particles
  diffusion at short time scales was systematically neglected. This regime is
  visible in Fig.~\ref{fig:monomerMSD}A for particle sizes $=30$ nm and $=60$
  nm.}\BibitemShut {Stop}%
\bibitem [{LTr()}]{LTransformNote}%
  \BibitemOpen
  \href@noop {} {}\bibinfo {note} {The Laplace transform,
  ${\tilde f}(s)$, of a function $f(t)$ with $t>0$ is defined as ${\tilde f}(s)
  \equiv \int_0^\infty f(t) \, e^{-st} dt$.}\BibitemShut {Stop}%
\end{thebibliography}


\end{document}